# CONVERTING ENERGY CAPTURED FROM BLOOD FLOW INTO USABLE ELECTRIC POWER: DESIGN OPTIMIZATION

Maria Angelika- Nikita[1,2], Neoclis Hadjigeorgiou[1], Christos Manopoulos[1] and Julius Georgiou[2]

[1]National Technical University of Athens, School of Mechanical Engineering, Iroon Polytechniou 9, Zografos 15780, Greece
[2]University of Cyprus, Department of Electrical and Computer Engineering, Panepistimiou 1, Aglantzia 2109, Nicosia, Cyprus

Email: Julio@ucy.ac.cy

**Abstract**
In this work, we attempt to optimize the design of an electromagnetic induction device that captures the energy from the arterial wall pulsation for the purpose of powering implantable biomedical devices. The artery comes through a flexible coil which is permitted to freely deform along with the artery in a magnetic field produced by two permanent ring magnets that are placed in parallel. As a result of the coil's motion in the magnetic field, an alternating voltage proportional to the velocity of the arterial wall is induced at the coil's terminals. The coil consists of a main loop which is aligned with the magnets' holes and a number of side loops fabricated using enameled copper wire 0.05 mm in diameter. In an attempt to increase the output power of the device, different coil geometries were developed with varying numbers and sizes of side loops, using 3D printed molds. An experimental setup that mimics blood flow and arterial wall deformation was used to assess the device performance. The acquired measurements demonstrated that the produced voltage and power can be notably increased, by increasing the number and diameter of the coil's side loops.

## 1. Introduction

On-body and in-body microsystems, like pacemakers, blood pressure sensors, drug-delivery pumps, glucose meters, or neurostimulators have the potential to revolutionize health care. Their wider adoption requires overcoming restrictions imposed by their source of power (Kim et al. 2015). Currently, external power sources or replaceable implanted batteries are used to provide the energy for wearable and implantable biomedical devices (Scrosati et al. 2013, Kim et al. 2015). However, batteries frequently dominate the size, and sometimes the cost of the devices, have a finite lifetime and chemical side effects, while their replacement requires surgery. Wirelessly powered devices, on the other hand, are battery-less but require an external interface for operation, which is burdensome and aesthetically unappealing (van Schuylebergh and Puers 2009).

Alternative power sources that overcome these limitations are, thus, highly desirable. Extracting power from ambient sources, known as energy harvesting or energy scavenging, constitutes one possible approach to this challenge. Although energy harvesting cannot produce the energy density of batteries, the vast reductions in power consumption achieved in electronics (Chandrakasan et al. 2008) are continuously increasing the attractiveness of harvesting techniques for on-body and in-body medical devices. Light, radio-frequency (RF) electromagnetic radiation, thermal gradients, and motion, including fluid flow, are essentially the forms of energy sources that are available for harvesting (Mitcheson et al. 2008). However, sources of ambient energy are limited for implanted microsystems, given that there are no consistent and substantial temperature gradients in the body, light penetration into the body is minimal and movement and vibration are not guaranteed (Mitcheson et al. 2008).

For this reason, researchers have been studying energy harvesting techniques directly from the human body. Living subjects have abundant sources of energy in chemical, thermal, and mechanical forms (Kim et al. 2011, Priya and Inman 2009). The use of these energy sources presents a viable way to eliminate the need for batteries and their shortcomings for implantable medical devices. The intersection of novel materials and fabrication techniques offers boundless possibilities for the benefit of human health and well-being via various types of energy harvesters. These devices harvest or recover a portion of the energy from body heat, breathing, arm motion, leg motion or the motion of other body parts produced during walking or any other activity (Cadei et al. 2014). In particular, kinetic energy harvesters are based on piezoelectric, electromagnetic and electrostatic conversion (Kim et al. 2011, Kim et al. 2015).

Early examples of piezoelectric energy capturing from body motion include the shoe energy harvester of the MIT Media Lab (Schenk and Paradiso 2001), which was followed by several other piezoelectric powered devices with applications ranging from blood pressure monitoring to orthopedic implants and knee replacements (Mitcheson et al. 2008, Kim et al. 2011). More recently, the technique has been adopted to capture energy from the motion of internal organs (Hwang et al. 2014) and an energy harvester based on thin PZT ribbons able to produce power density per PZT area 1.2 W/cm$^2$ from heart motion was fabricated (Dagdeviren et al. 2014). Moreover, energy harvesting from arterial blood pressure for powering embedded micro-sensors in the human brain has been proposed, achieving a peak harvested power of 11.7 $\mu$W for the carotid artery and a predicted peak power of 203.4 $\mu$W for the aorta (Nanda and Karami 2017).

Apart from piezoelectric techniques, electromagnetic induction principles have been applied to capture the energy from the low-frequency and irregular movements of humans and transform it into low power electricity for powering medical implants. An electromagnetic device 16 cm$^3$ in volume, able to generate about 1.1 mW from the abdomen during breathing at a frequency of 0.3 Hz has been developed (Ben Amar et al. 2015). In addition, several efforts have been reported towards the development of energy harvesting methods, which exploit motions of the cardiovascular system to produce electrical energy through electromagnetic induction. With the periodic contraction of the heart muscle at a frequency of 0.5-2 Hz, a power of 40–200 µW has been generated (Irani et al. 2009), while an electromagnetic generator based on MEMS technology has been developed to enhance the power for pacemaker batteries in clinical trials (Roberts et al. 2008). A generator based on a novel endocardial energy harvesting concept has been recently proposed (Zurbuchen et al. 2018), able to produce a mean output power of 0.78 and 1.7 µW at a heart rate of 84 and 160 beats per minute (bpm), respectively. Moreover, an electromagnetic mechanism that generated 42 nW from arterial wall deformation has been constructed (Pfenniger et al 2013). Although implementations of energy harvesters are showing progress on miniaturization and practical MEMS devices are beginning to appear, reported power levels remain well below theoretical maxima (Kim et al. 2015).

In this work, we attempt to optimize the design of an electromagnetic induction device that captures the energy from the arterial wall pulsation. A flexible coil is placed around the artery and is permitted to freely deform along with the artery in a magnetic field produced by two permanent ring magnets that are placed in parallel. Aiming at increasing the output power of the device, different coil geometries were fabricated and tested using an experimental setup that mimics blood flow and arterial wall deformation at different frequencies, corresponding to different heart rates and subject's states.

## 2. Methods

### 2.1 Working Principle of the Energy Harvester

The device under study consists of two permanent ring magnets placed in parallel. A flexible coil constructed with side loops, as depicted in Figure 1, is placed between the magnets, aligned with the magnets' holes. The artery comes through the holes of the magnets and the main loop of the coil. The magnets are magnetized along the artery axis. The coil is permitted to freely deform along with the artery inside the magnetic field produced by the magnets, and thus, an alternating voltage proportional to the velocity of the arterial wall is induced at the coil's terminals. Since the side loops move inside the magnetic field, they contribute to the induced voltage. Moreover, the presence of the side loops makes the coil flexible, allowing it to deform freely without restraining arterial wall movement. For the same reason, the coil's thickness should be constrained so that it does not modify arterial wall dynamics. In an attempt to optimize the efficiency of the energy harvesting device, different coil geometries were investigated consisting of a main loop with a fixed diameter and three or four side loops of varying diameter (Figure 2(b)).

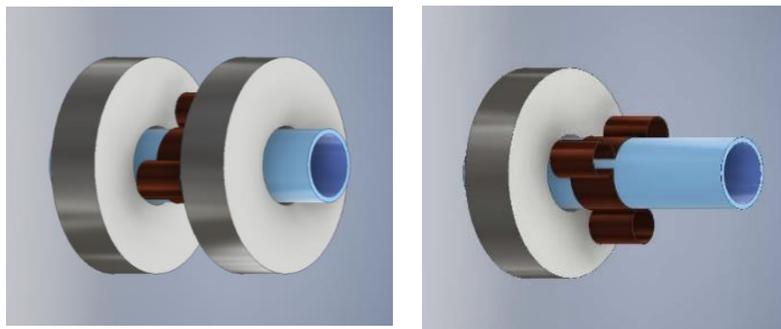

Figure 1: The harvesting device under study consists of two ring magnets and a flexible coil. The effect of the number and diameter of the coil's side loops on output voltage is examined

## 2.2 Theoretical Background

Figure 2 shows the ring magnets and the coordinates system. The artery extends along the z-axis and its wall deforms along the radial direction $r$. The two magnets have a length $l$, inner and outer radius $R_{in}$ and $R_{out}$, respectively, remanence $B_r$ and they are separated by a distance $s$. Given that the artery deforms radially, only the z-component of the magnetic field $B_z(r, z)$ contributes to the voltage induction.

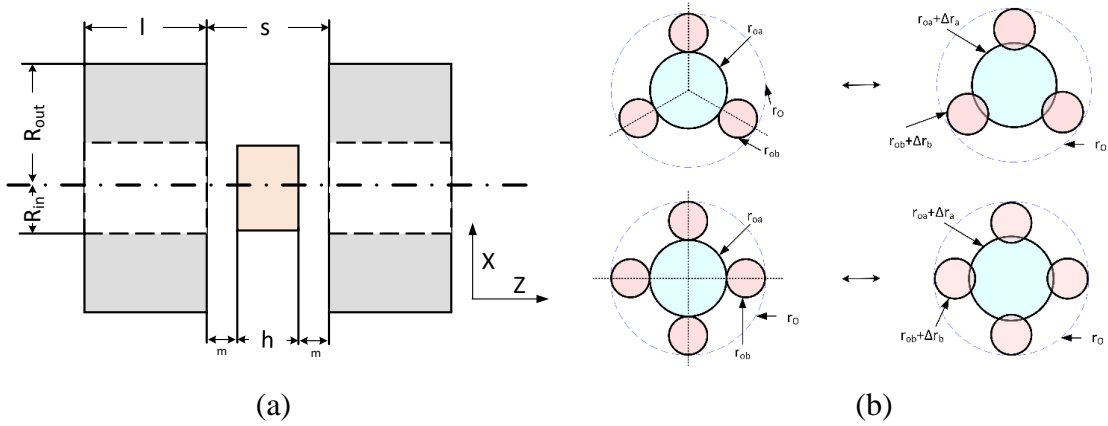

(a) (b)

Figure 2: (a) Ring magnets and coordinates system, (b) coil structures with three or four side loops at undeformed (left) and deformed (right) position

We assume a sinusoidal deformation of the outer artery wall described by the equations,

$$r_a(t) = r_{0a} + \Delta r_a(t), \text{ with } \Delta r_a(t) = \frac{\Delta r_a}{2}(1 - \cos(2\pi f t)) \tag{1}$$

where $r_{0a}$ is the artery's undeformed outer radius, $\Delta r_a$ is the maximum radial deformation of the artery, and $f$ is the frequency of the pressure pulse. The radial position $r_a(t)$ of the main loop is given by eq. (1), while the position $r_b(t)$ of the side loops is given by

$$r_b(t) = r_{0b} + \Delta r_b(r_a(t), t) \tag{2}$$

where $r_{ob}$ is the undeformed radius of the side loops and $\Delta r_b(t)$ is the deformation of the side loops which is obtained through simple geometrical calculations from $\Delta r_a(t)$.

The arterial wall and coil velocity is given by

$$v(t) = \frac{dr_a(t)}{dt} = \pi f \Delta r_a \sin(2\pi f t) \tag{3}$$

The total resistance ($R_c$) and inductance ($L_c$) of each coil are computed as,

$$R_c = \rho_w \frac{l_w}{A_w} = \rho_w \frac{8N(r_{0a} + nr_{0b})}{d_w^2} \tag{4}$$

$$L_c = \frac{\mu_0 N^2}{\pi h}(r_{0a}^2 + nr_{0b}^2) \tag{5}$$

where $N$ is the number of windings that form the coil, $\rho_w$, $l_w$, $A_w$ and $d_w$ are the resistivity, length, surface and diameter of the wire, respectively, $\mu_0$ is the air magnetic permeability, $h$ is the height of the coil and $n$ is the number of coil's side loops.

The induced voltage at the terminals of the coil is

$$V_{out}(t) = -L_c \frac{d(B_z A(t))}{dt} \tag{6}$$

where $A(t)$ is the active area of the coil which is computed through simple geometric manipulations from Figure 2(b). The power transfer of the device is maximum when the coil is connected in parallel with a resistance equal to its own and can be expressed as,

$$P_{max}(t) = \frac{(V_{out}(t)/2)^2}{R_c} \tag{7}$$

The magnetic field produced by the ring magnets, as well as the open circuit output voltage and power with respect to arterial wall displacement are computed using Finite Element Analysis by means of the ANSYS Maxwell electromagnetic field simulation software (Ansys 2018). To define the spacing between the magnets that maximizes the output power, a parametric study was carried out, where $\Delta r_a/r_a$ was set at 10%, corresponding to a typical increase in vascular diameter between diastole and systole. A coil with $h$=10 mm, consisting of a main loop $2r_a$=14 mm in diameter, made of $N$=590 turns from enameled copper wire ($\rho_w$=1.68x10$^{-8}$ Ωm) with a diameter $d_w$=0.05 mm, wound in four layers was considered, placed between two ring magnets with $l$=10 mm, $R_{in}$=7.5 mm, $R_{out}$=20.5 mm, and $B_r$=1.2 T (Figure 2). Second order triangular elements 0.05 mm in size were used for coil meshing and adaptive meshing was used for the air and magnets. The output power as a function of spacing ($s$) is presented in Figure 3, where it can be seen that the peak harvested power is obtained for $s$=20 mm. In Figure 4, the corresponding distribution of the longitudinal magnetic field magnitude ($B_z$) is shown for $f$=1.25 Hz (HR=75 bpm).

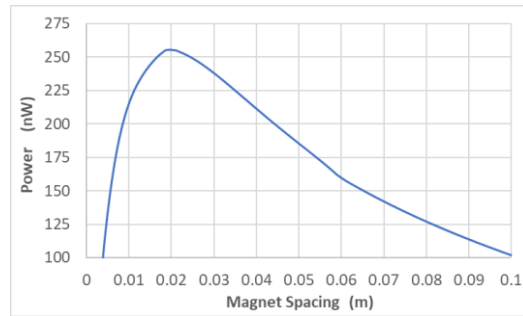

Figure 3: Generated power as a function of the spacing between the magnets

**2.3 Coil Fabrication**

Two axially magnetized neodymium ring magnets with $B_r$=1.2 T, $l$=10 mm, $R_{in}$=7.5 mm and $R_{out}$=20.5 mm were placed at a constant spacing ($s$=20 mm) through non-magnetic, stainless steel screws. Different coil geometries with a fixed total number of $N$=590 turns were constructed using enameled copper wire ($\rho_w$=1.68x10$^{-8}$ Ωm) with a diameter $d_w$=0.05 mm, wound in four layers. This led to a surrounding circular structure of up to 2 mm in thickness, which seems acceptable for mid- to large-sized arteries, as it is not considered to modify the arterial wall dynamics. The studied coil geometries consisted of a main loop with fixed diameter ($2r_a$ = 14 mm) surrounded by either (a) three side loops with individual diameter $2r_b$=7 mm (coil 3P7), or (b) three side loops with $2r_b$= 10 mm (coil 3P10), or (c) four side loops with $2r_b$= 7 mm (coil 4P7) or (d) four side loops with $2r_b$=10 mm (coil 4P10). The above coils were fabricated using 3D printed winding and shaping molds from polyvinyl alcohol (PVA), a water-soluble support material for multi-extrusion 3D printing. The complete fabrication procedure is shown in Figure 5. The measured and estimated resistance and inductance values of the different coils exhibit satisfactory agreement, as shown in Table 1. Observed deviations are attributed to potential inaccuracies in the coil fabrication procedure and underestimation of measured inductance due to parasitic capacitance.

**2.4 Experimental setup**

The device performance was tested using an experimental setup that mimics blood flow and arterial wall deformation (Figure 6). An elastic tube made of platinum cured silicon of 12 mm inner diameter and 1 mm wall thickness was used to simulate the artery. In order to perform pressure measurements, a tube system was formed by connecting the elastic tube

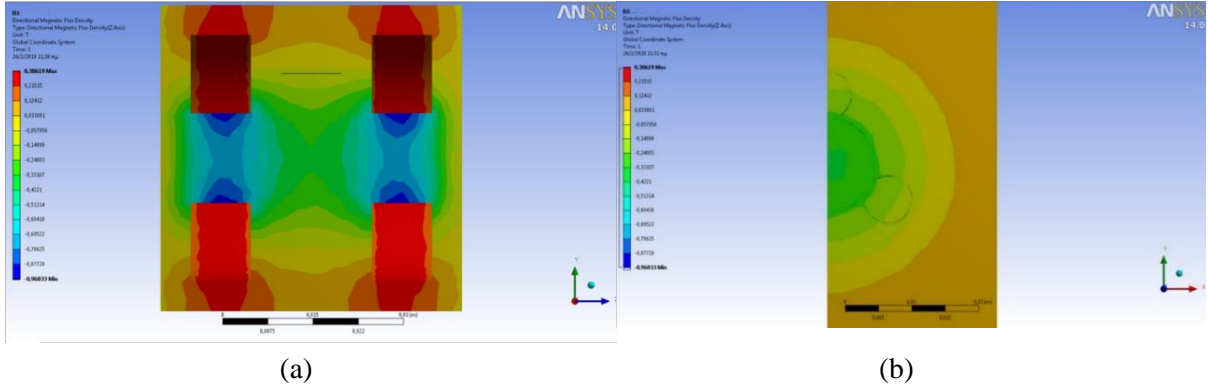

(a)                                                                   (b)

Figure 4: Distribution of the magnetic field magnitude $B_z$ produced by the ring magnets on the medial (a) yz and (b) xy planes

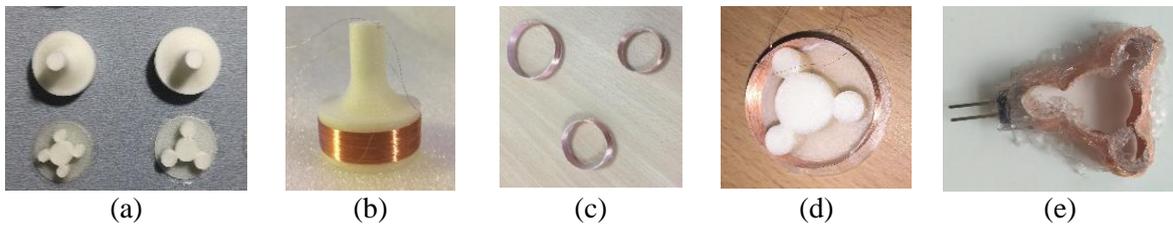

(a)             (b)             (c)             (d)             (e)

Figure 5: Coil fabrication. (a) 3D printed winding and shaping molds from PVA, (b) coil wound around the winding mold, (c) coils after PVA winding support removal by dissolving in warm water, (d) coil placed in the shape forming mold, (e) coil after removal of the shaping mold

Table 1: Estimated and measured values of the resistance and inductance for the four coils

|  | **Resistance $R_c$ (Ω)** | | | | **Inductance $L_c$ (mH)** | | | |
|---|---|---|---|---|---|---|---|---|
| **Coil** | 3P7 | 3P10 | 4P7 | 4P10 | 3P7 | 3P10 | 4P7 | 4P10 |
| **Estimated** | 482 | 598 | 624 | 757 | 11.7 | 17.0 | 13.4 | 19.4 |
| **Measured** | 525 | 685 | 743 | 874 | 8.1 | 11.4 | 9.1 | 12.7 |

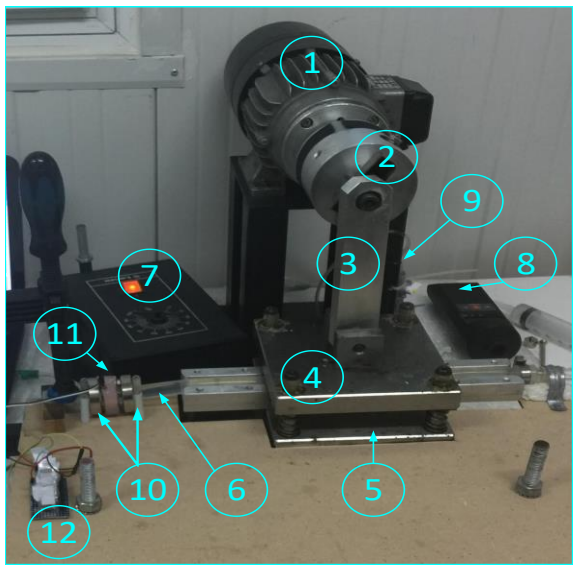

The energy harvesting device consists of two ring magnets (10) and the pickup coil (11). The elastic silicon tube (6) comes through the energy harvesting device and is connected to a rigid tube where the catheter pressure sensor (9) is inserted. A pressure pulse is applied to a segment of the elastic tube by a device consisting of an electric motor (1) with a flywheel (2), a piston rod (3), a piston – plate (4), and a support plate (5). A voltage control unit (7) is used to adjust the motor rotation speed while a rotation speed meter (8) is used to measure the frequency of the produced pressure pulse. The time waveforms of the induced voltage are measured for different load resistances (load matching array block (12)) to a rigid tube of the same radius made of Plexiglas and a catheter was inserted to the latter through a thin hole on its wall. The other end of the catheter was connected to an Edwards Truwave pressure disposable transducer integrated in a pressure monitoring kit (Edwards Life Sciences, 2018). The tube system was filled with water and all the air was removed through a properly connected valve. A motor-based device was used to compress a segment of the elastic tube periodically, causing a periodic pressure pulse inside the tube with adjustable pressure and frequency. The temporal waveforms of the pressure inside the tube system and the output voltage at the terminals of the coil were synchronously sampled by using a HBM Quantum X MX440A Analog to Digital Converter.

Figure 6: Experimental setup

## 3. Results and Discussion

The variation of the harvested power with parameters of the coil design, i.e. number and size of the side loops, and pulse rate was investigated. Measurements were acquired for coils 3P7, 3P10, 4P7 and 4P10 with varying load resistances (1.1, 10, 51, 110, 220, 300, 430, 510, 620, 750, 820, 910, 1.2K,1.6K, 2.2K, 3K, 3.9K Ω and open circuit). The induced voltage was measured at four different frequencies corresponding to heart rates (HR) of 52, 75, 110, 135 ±2 bpm that are related to different subject's states (sleep, normal, tachycardia and exercise). A pressure pulse of amplitude 284.42 mmHg was applied inside the elastic tube producing a 10% deformation. The output power was estimated from Eq. (7) through voltage measurements, for loads matching each coil's resistance, whereby the power transfer is maximum. Numerical results were obtained using the ANSYS simulation software (Ansys 2018) with second order triangular elements 0.05 mm in size for coil meshing and adaptive meshing for the air and magnets, leading to a total number of 2.75 M, 3.2 M, 3.455 and 4.24 M elements for coils 3P7, 3P10, 4P7, 4P10, respectively.

In Figure 7, the measured output voltage and power as a function of the load resistance are shown for each of the coils. Their variation with HR is also presented. The maximum output power is observed at load resistance close to the coil's resistance, while the output voltage and power increase with increasing HR, as expected.

In Figure 8, the output voltage measured across the coil's terminals and the maximum generated power is shown at each frequency for the coils under study. By increasing the diameter of the side loops, a consistent increase in the induced voltage is observed, as expected, which is more pronounced for the case of the small-diameter loops at higher frequencies. Given that the side loops deform along with the artery, they contribute to the voltage induction. Hence, the induced voltage is expected to increase with increasing number of side loops. This is in agreement with measurements acquired for the small-sized loops. However, the voltage induced in the case of the coil with four side loops 10 mm in diameter (4P10) is lower than that of the coil with three side loops (3P10). This was attributed to the remaining deformation of the coil 4P10, which did not allow it to closely follow the entire range of arterial wall deformation. To further investigate this deviation from the theoretically expected results, an identical coil with four side loops 10 mm in diameter was fabricated and a slightly stronger silicone support (4P10Sil) was used to avoid permanent deformation and allow the main loop to maintain its contact with the external vascular wall during deformation. The output voltage and power generated by the new coil are higher than their corresponding values for coil 3P10 (Figure 8), as expected. This observation indicated the importance of fabricating coils with elastic shape properties able to follow the arterial wall deformation during expansion and contraction of the vessel.

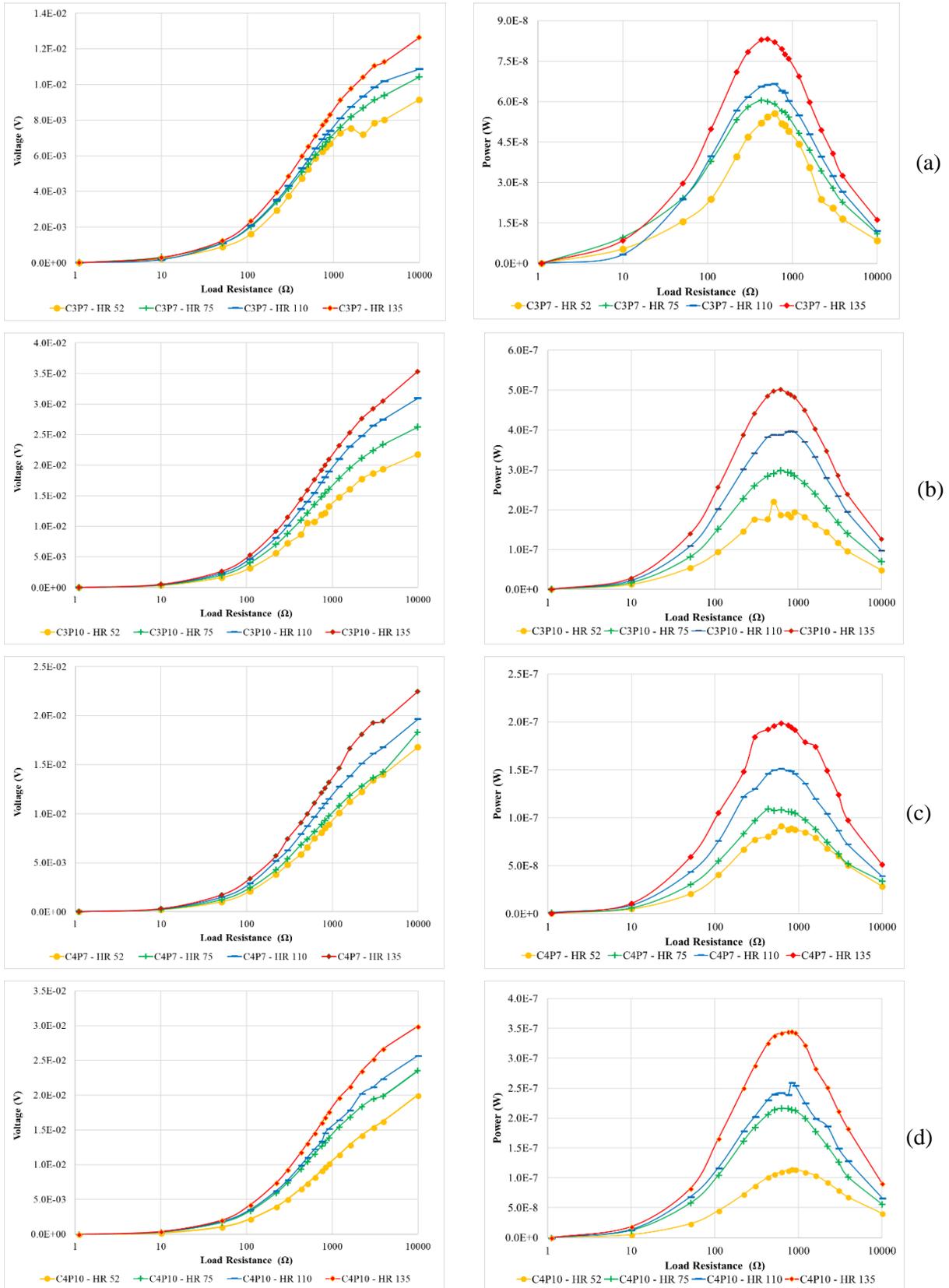

Figure 7: Measured output voltage and power as a function of load resistance at different heart rates (HR) for (a) coil 3P7, (b) coil 3P10, (c) coil 4P7, (d) coil 4P10

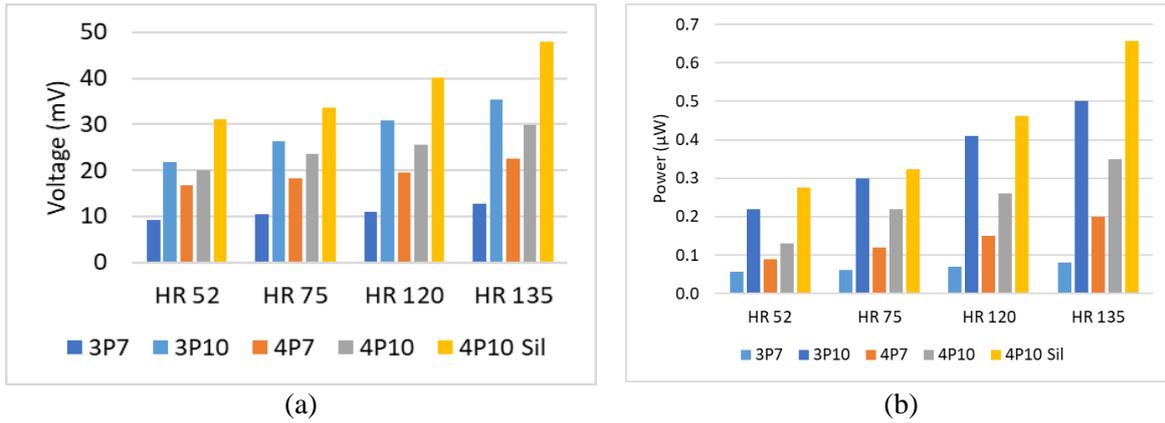

(a)                                      (b)

Figure 8: Induced voltage and maximum output power for each coil at different heart rates

In Figure 9, the induced voltage and harvested power are shown for the coils with three or four side loops 10 mm in diameter with increased silicone support (3P10 and 4P10Sil) at different frequencies. Measurements are presented along with computations obtained using the ANSYS Maxwell software package, by considering either only the main loop of the coil (S-Coil) or the full coil structure with three or four side loops (F-Coil). It can be observed that the experimental value is always between the one corresponding to the coil consisting of the main loop only and that corresponding to the full coil structure. This means that the fabricated coil's performance is improved with the addition of the side loops, however it does not achieve the theoretically predicted performance, due to fabrication flaws, e.g. the coil suffers remaining deformation that does not allow it to closely follow the arterial wall's radial movement. The deviation from the theoretically predicted performance for the full coil structure is more pronounced for higher heart rates.

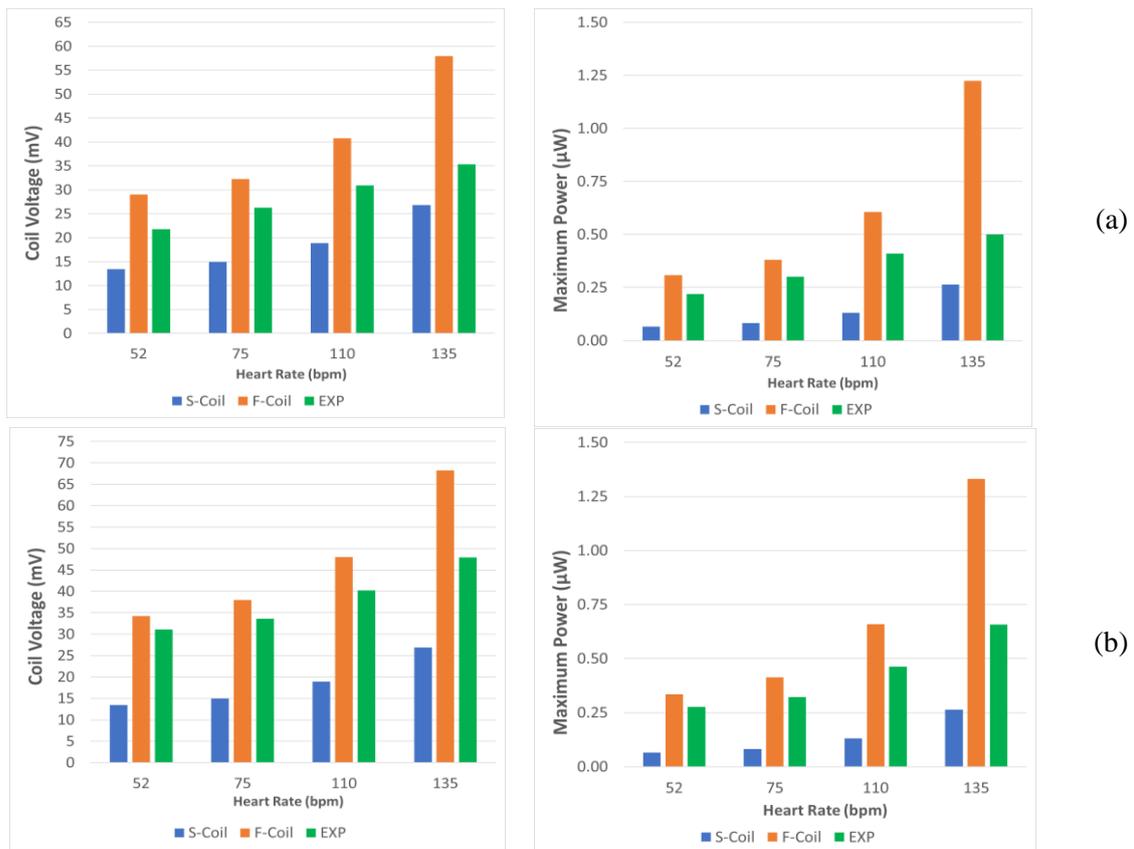

Figure 9: Induced voltage and maximum output power for (a) coil 3P10 and (b) coil 4P10Sil, at different heart rates (experimental values (EXP), simulated values for a coil consisting of the main loop only (S-Coil) or for the complete coil structure (F-Coil))

By using subthreshold circuit design techniques (Georgiou and Toumazou, 2005), (Constandinou et al. 2003) and/or current mode circuits, power levels ranging between 100 nW to 500 nW can easily power an implanted amplifier, a reference circuit and an analog filter. In its current form, the output voltage is barely sufficient for using transistor based power management units (Bandyopadhyaay et al., 2014). However, this can be optimized in a future version by using more windings along with thinner diameter coils (e.g. 15 μm silver) or by applying techniques involving a piezoelectric transformer (Martinez et al., 2016) to boost the low voltages to much more suitable levels.

4. Conclusions

Numerical simulations and experimental measurements of the electromagnetic energy harvester under study demonstrated that a notable increase in the induced voltage and harvested power can be achieved by increasing the number of side loops and/or their individual diameter. However, special care should be taken to ensure proper flexibility of the coil so that it does not constrain the arterial wall movement and at the same time maintains its elastic shape properties and ability to closely follow the arterial wall deformation during expansion and contraction of the blood vessel.